\documentclass[%
 aip,
 amsmath,amssymb,
reprint,%
]{revtex4-1}

\usepackage{graphicx}% Include figure files
\usepackage{dcolumn}% Align table columns on decimal point
\usepackage{bm}% bold math
\usepackage{setspace}
\usepackage[T1]{fontenc}
\usepackage{mathptmx}
\usepackage{etoolbox}
\usepackage{float}
\usepackage{tabularx}
\usepackage{xcolor}
\usepackage{outlines}

\usepackage{amsmath,amssymb}

\usepackage{appendix}

\usepackage{xr}
\makeatletter
\newcommand*{\addFileDependency}[1]{% argument=file name and extension
  \typeout{(#1)}
  \@addtofilelist{#1}
  \IfFileExists{#1}{}{\typeout{No file #1.}}
}
\makeatother
\newcommand*{\myexternaldocument}[1]{%
    \externaldocument{#1}%
    \addFileDependency{#1.tex}%
    \addFileDependency{#1.aux}%
}
\myexternaldocument{SI}

\usepackage{appendix}

\begin{document}

\singlespacing

\title{Tailoring interactions between active nematic defects with reinforcement learning}
% Force line breaks with \\
\author{Carlos Floyd}
\email{csfloyd@uchicago.edu}
% \affiliation{The Chicago Center for Theoretical Chemistry, The University of Chicago, Chicago, Illinois 60637, USA
% }
\affiliation{Department of Chemistry, The University of Chicago, Chicago, Illinois 60637, USA
}
\affiliation{The James Franck Institute, The University of Chicago, Chicago, Illinois 60637, USA
}
\author{Aaron R.\ Dinner}
% \affiliation{The Chicago Center for Theoretical Chemistry, The University of Chicago, Chicago, Illinois 60637, USA
% }
\affiliation{Department of Chemistry, The University of Chicago, Chicago, Illinois 60637, USA
}
\affiliation{The James Franck Institute, The University of Chicago, Chicago, Illinois 60637, USA
}
\author{Suriyanarayanan Vaikuntanathan}
\email{svaikunt@uchicago.edu}
% \affiliation{The Chicago Center for Theoretical Chemistry, The University of Chicago, Chicago, Illinois 60637, USA
% }
\affiliation{Department of Chemistry, The University of Chicago, Chicago, Illinois 60637, USA
}
\affiliation{The James Franck Institute, The University of Chicago, Chicago, Illinois 60637, USA
}

\date{\today}% It is always \today, today,

\begin{abstract}
Active nematics, formed from a liquid crystalline suspension of active force dipoles, are a paradigmatic active matter system whose study provides insights into how chemical driving produces the cellular mechanical forces essential for life.  Recent advances in optogenetic control over molecular motors and cell-signaling pathways now allow experimenters to mimic the spatiotemporal regulation of activity necessary to drive biologically relevant active nematic flows \textit{in vivo}.  However, engineering effective activity protocols remains challenging due to the system's complex dynamics. Here, we explore a model-free approach for controlling active nematic fields using reinforcement learning. Specifically, we demonstrate how local activity fields can induce interactions between pairs of nematic defects, enabling them to follow designer dynamical laws such as those of overdamped springs with varying stiffnesses.  Reinforcement learning bypasses the need for accurate parameterization and model representation of the nematic system, and could thus transfer straightforwardly to experimental implementation.  Moreover, the sufficiency of our low-dimensional system observables and actions suggests that coarse projections of the active nematic field can be used for precise feedback control, making the biological implementation of such feedback loops plausible.
\end{abstract}

\maketitle
\begin{twocolumngrid}

\section{Introduction}
To produce physiologically useful forces, the cytoskeletal machinery of cells must be tightly regulated and controlled \cite{fletcher2010cell, banerjee2020actin, levine2023physics}.  For example, to execute cell division the cytoskeleton must localize to the midplane of cells and pinch in a highly coordinated event \cite{barr2007cytokinesis}.  During morphogenesis the control of mechanical forces must be coordinated at the tissue level, across many different cells.  It has recently been shown that during this process defects in the nematic ordering of cytoskeletal filaments in epithelial cells of developing \emph{Hydra} are precisely positioned at key global organizing centers, such as the future mouth \cite{maroudas2021topological}.  Related research indicates that the positioning of nematic defects plays a crucial role in organizing the stress and velocity fields of the system, driving large-scale coherent motions of the tissue layer \cite{guillamat2022integer, serra2023defect, shankar2022topological}.  Although the control of cytoskeletal machinery is known to involve factors like cell signaling, mechanical interactions with the environment, chemical activation of cytoskeletal proteins, and mechanochemical feedback loops, how these various factors coordinate to produce coherent and functional cytoskeletal forces which sculpt tissue layers remains unclear \cite{levine2023physics,bruckner2024tissue, mitchell2022visceral}.

One approach to addressing this question is to map biochemical interactions among known molecular players to identify potential control loops. For example, recent works have shed light on the origins of contractile oscillations during cytokinesis in \textit{C. elegans} and used mechanochemical feedback loops to explain polarity establishment during morphogenesis \cite{werner2024mechanical, maxian2024minimal}. An alternative strategy is to work from the top down: without detailed biochemical knowledge, we can explore control protocols that dynamically adjust downstream cytoskeletal inputs to achieve specific system motions, offering insights to guide the search for underlying molecular mechanisms.  This approach has been used in recent work on active nematics by directly controlling spatiotemporally dynamic activity fields $\alpha(\mathbf{r},t)$, viewed as a downstream outcome of unresolved biochemical circuits. Experimental advances in spatiotemporal motor control using light fields provide an additional and practical reason for studying activity fields as externally controlled functions \cite{linsmeier2016disordered, schindler2014engineering, nakamura2014remote, zhang2021spatiotemporal, lemma2023spatio, chandrasekar2023shining}. In simulations, optimal activity field trajectories $\alpha(\mathbf{r},t)$ can be derived using knowledge of the nematohydrodynamic equations of motion to guide nematic and polar defects along desired paths \cite{norton2020optimal, ghosh2024spatiotemporal, ghosh2024achieving}. Other techniques allow targeted modulation of nematic channel flow and design of localized "topological tweezers" for precise defect manipulation \cite{wagner2022exact, shankar2024design, irvine2013dislocation}.

A key challenge in implementing the above mentioned control methods is their reliance on accurate system models and parameterization.  Here, we explore controlling active nematics through a model-free machine learning technique called reinforcement learning (RL) \cite{sutton2018reinforcement, bechhoefer2021control}. This approach allows a program to develop a closed-loop control policy for a dynamical system purely through trial and error, without relying on precise model specifications.  To our knowledge RL has not yet been applied to control active nematics, although previous studies have used it to control other active matter systems such as active flockers, active crystallization, and branched actin networks \cite{falk2021learning, chennakesavalu2021probing, chennakesavalu2024adaptive}. These studies focus on design tasks that aim to achieve a specific static property of the system, such as target net flocking motion or average crystal size. In contrast, we study control tasks that guide active nematic defects to follow prescribed virtual interactions with one another (see Figure \ref{SchematicOverview}) \cite{khadka2018active, bauerle2018self}. Our goal is to learn activity field protocols that can effectively override the natural defect dynamics (such as passive Coulomb-like attraction) to impose a user-specified interaction law, a more general design objective termed `cyberphysics' in Ref.~\citenum{bechhoefer2021control}.  Additionally, we constrain the program to learn imperfect, yet efficient, protocols by only allowing it to view a coarse projection of the full nematic field configuration for its feedback control. This approach suggests that techniques for controlling cytoskeletal materials can be effective using simple and approximate feedback loops, which has implications for refining experimental control over active nematics using light-controlled motors.  Relatedly, in previous work we demonstrated that dynamic activity fields can be iteratively constructed without RL and without knowledge of the active nematic dynamics using physically motivated yet imperfect local feedback rules \cite{floyd2024learning}.  The sufficiency of these imperfect feedback control techniques can provide insights into the components necessary for mechanochemical feedback loops to regulate cytoskeletal machinery in cells.

\begin{figure}[ht!]
\begin{center}
\includegraphics[width=\columnwidth]{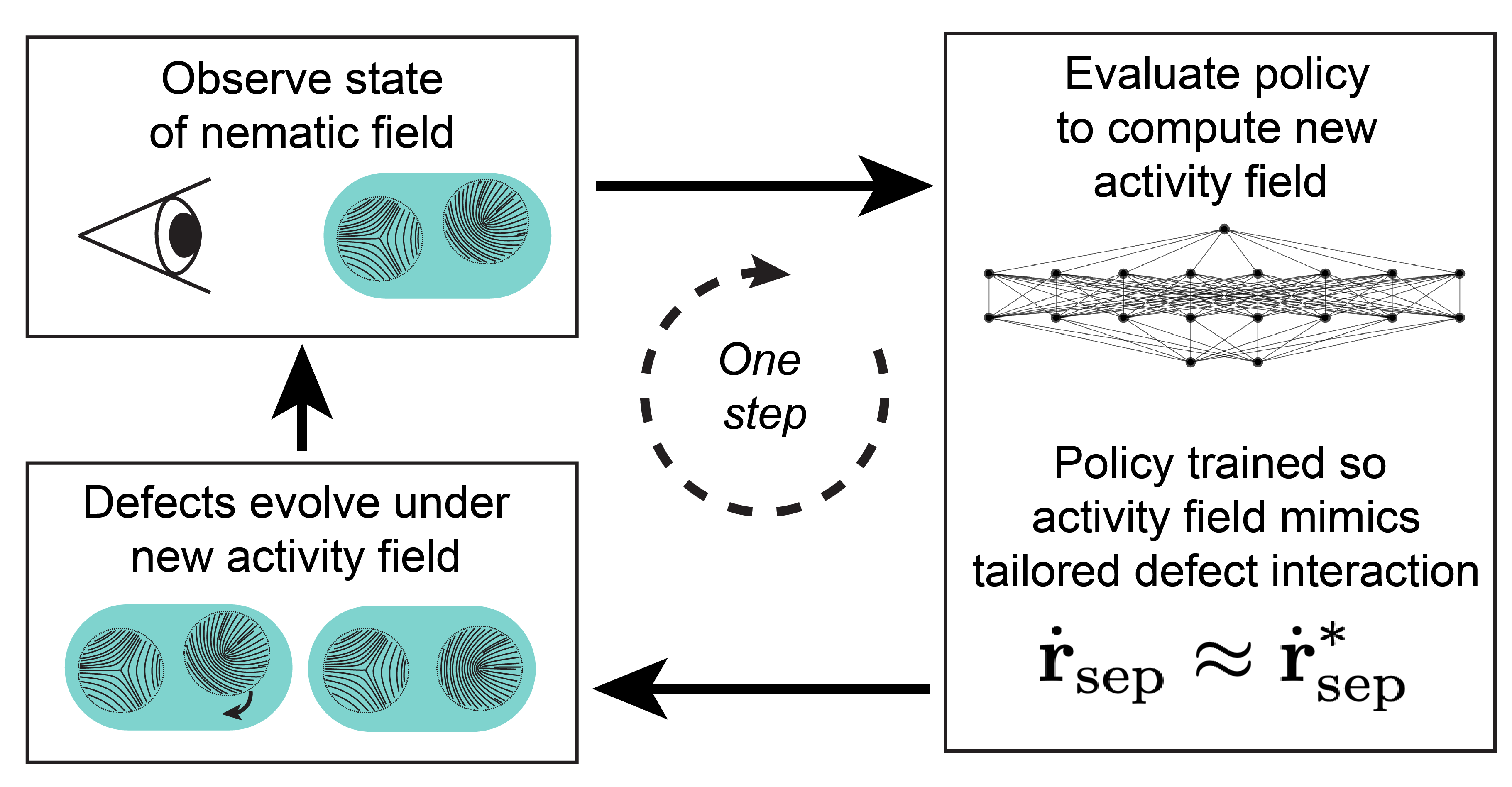}
\caption{Schematic overview of the closed-loop control over active nematic defect dynamics enabled by a trained RL policy.  One step of the feedback loop is depicted (see Figure \ref{TimeDivisions}).}
\label{SchematicOverview}
\end{center}
\end{figure}

\section{Methods}
Here we describe how we formulate and solve the RL task for imposing customized interaction laws between active nematic defects.  In the Appendix we outline the overdamped active nematohydrodynamic equations of motion we simulate.  In Section \ref{sec:dgi} we describe the geometry and interaction laws governing active nematic defects.  We then define the states, actions, rewards, and experiment structure which comprise our RL setting in Section \ref{sec:sar}.  In Section \ref{sec:rl} we describe the specific RL algorithm we use to optimize the policy.  

\begin{figure}[ht!]
\begin{center}
\includegraphics[width=\columnwidth]{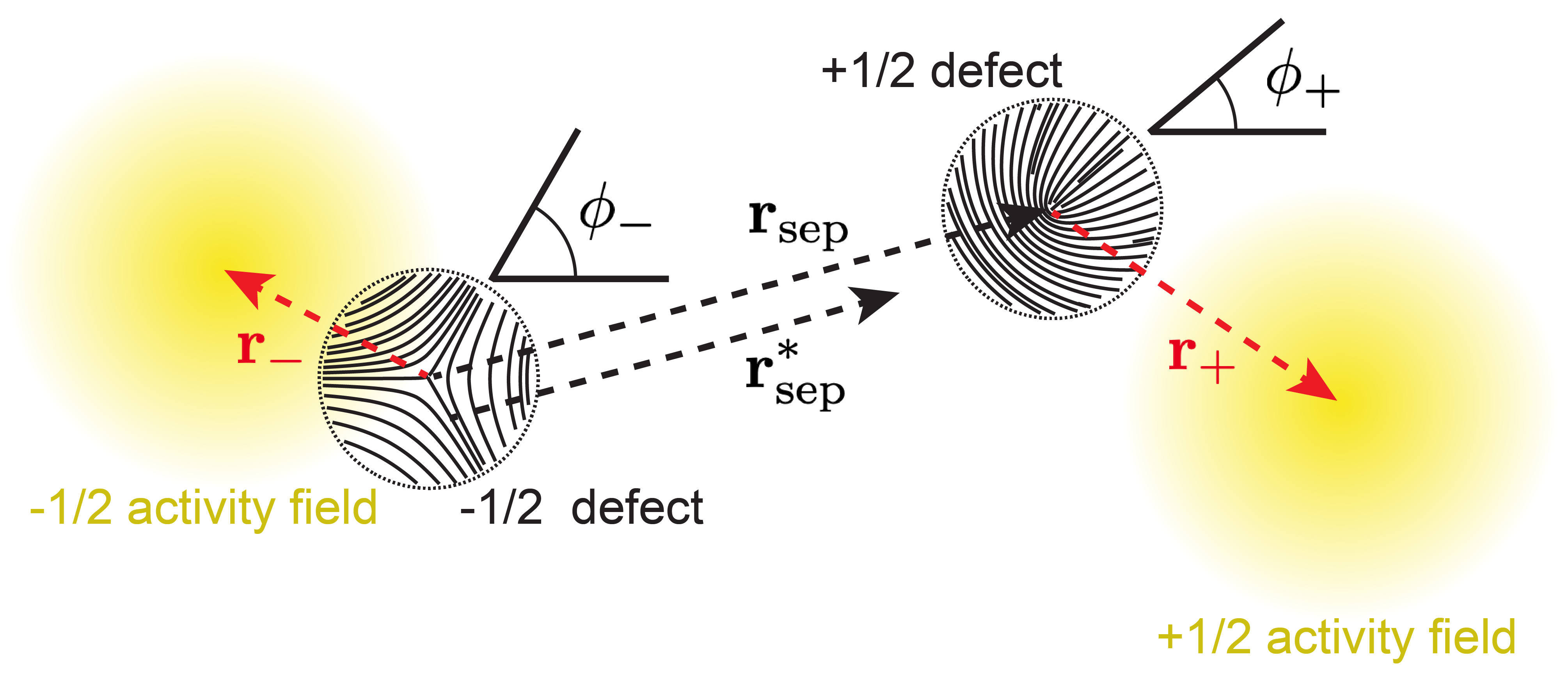}
\caption{Illustration of the geometric quantities used to define the positions and orientations of a pair of $\pm1/2$ nematic defects in a general configuration.  Two activity fields are also shown in the vicinity of the defects.  The vectors $\mathbf{r}_\pm$ point from the defect locations to the center of their nearby activity fields, while $\mathbf{r}_\text{sep}$ points from the $-1/2$ defect to the $+1/2$ defect.  The target separation $\mathbf{r}^*_\text{sep}$ is denoted with an asterisk.  The angles $\phi_\pm$ describe the orientation of the defects with respect to the horizontal axis.  }
\label{DefectGeometry}
\end{center}
\end{figure}

\subsection{Defect geometry and interactions}\label{sec:dgi}

We consider throughout a pair of $\pm 1/2$ defects in a periodic 2D domain. 
The defect positions are labeled $\mathbf{p}_{\pm}$ and we define the separation vector $\mathbf{r}_\text{sep} = \mathbf{p}_+ - \mathbf{p}_-$ (see Figure \ref{DefectGeometry}).  As described for example in Ref.~\citenum{shankar2024design}, each defect has an orientation in addition to its position.  The $+1/2$ defect orientation is described by a vector $\hat{\mathbf{e}} = (\cos(\phi_+), \sin(\phi_+))$ given by
\begin{equation}
    \hat{\mathbf{e}} = \frac{\nabla \cdot \mathbf{Q}(\mathbf{r},t)}{\left|\nabla \cdot \mathbf{Q}(\mathbf{r},t) \right|} 
\end{equation}
where $\mathbf{Q}$ is the symmetric and traceless nematic order parameter, and the expression is evaluated in the limit approaching the center of the defect core.  The orientation of the $-1/2$ defect has a three-fold symmetry and cannot be represented by a vectorial quantity.  Instead, it is represented by the rank-three tensor
\begin{equation}
    \Theta_{ijk} = \frac{\left\langle\partial_i Q_{jk} + \partial_jQ_{ik} + \partial_kQ_{ij} \right\rangle}{3\left|\left\langle \partial_kQ_{ij}\right\rangle \right|}
\end{equation}
where the brackets denote an angular average around the defect core.  This quantity can also be expressed in terms of triple outer products of a vector $\hat{\mathbf{t}} = (\sin(\phi_-), \cos(\phi_-))$, as
\begin{equation}
    \Theta_{ijk} = \hat{t}_i\hat{t}_j\hat{t}_k - \frac{1}{4}\left(\delta_{ij}\hat{t}_k + \delta_{kj}\hat{t}_i + \delta_{ik}\hat{t}_j \right)
\end{equation}
where $\delta_{ij}$ is the Kronecker delta.  As required, $\Theta_{ijk}$ is invariant under $\phi_- + n 2\pi/3$ for any integer $n$.  In simulation we compute the nematic positions and orientations following methods described in Ref. \citenum{tang2017orientation}.

Although defects in an active nematic fluid evolve under complex hydrodynamics (see Equations \ref{eqQdyn} and \ref{eqv} below), theoretical work has shown that their motion can be approximated using simpler one and two-body dynamical equations \cite{shankar2018defect, shankar2024design}.  The positions $\mathbf{p}_\pm$ and orientations $\phi_\pm$ approximately obey differential equations which are coupled to each other through elastic interactions, and they are also coupled to the local activity profile $\alpha(\mathbf{r})$ and its gradients.  These equations are derived from the hydrodynamic Stokes flow of the active nematic fluid and involve numerical prefactors such as the effective defect size, friction coefficient, elasticity constant, shear viscosity, and others.  These effective dynamical equations enable the design of precise ``activity tweezers'' that manipulate defect dynamics, but this technique is complicated by the difficulty of estimating the model parameters involved \cite{shankar2024design, joshi2022data}.  Here, we simply use the fact that defect positions and orientations couple in some way to activity gradients to find tweezer-like activity protocols purely through exploration (i.e., RL).  This approach is thus agnostic to the underlying model parameters and dynamics.   

\begin{figure}[ht!]
\begin{center}
\includegraphics[width=\columnwidth]{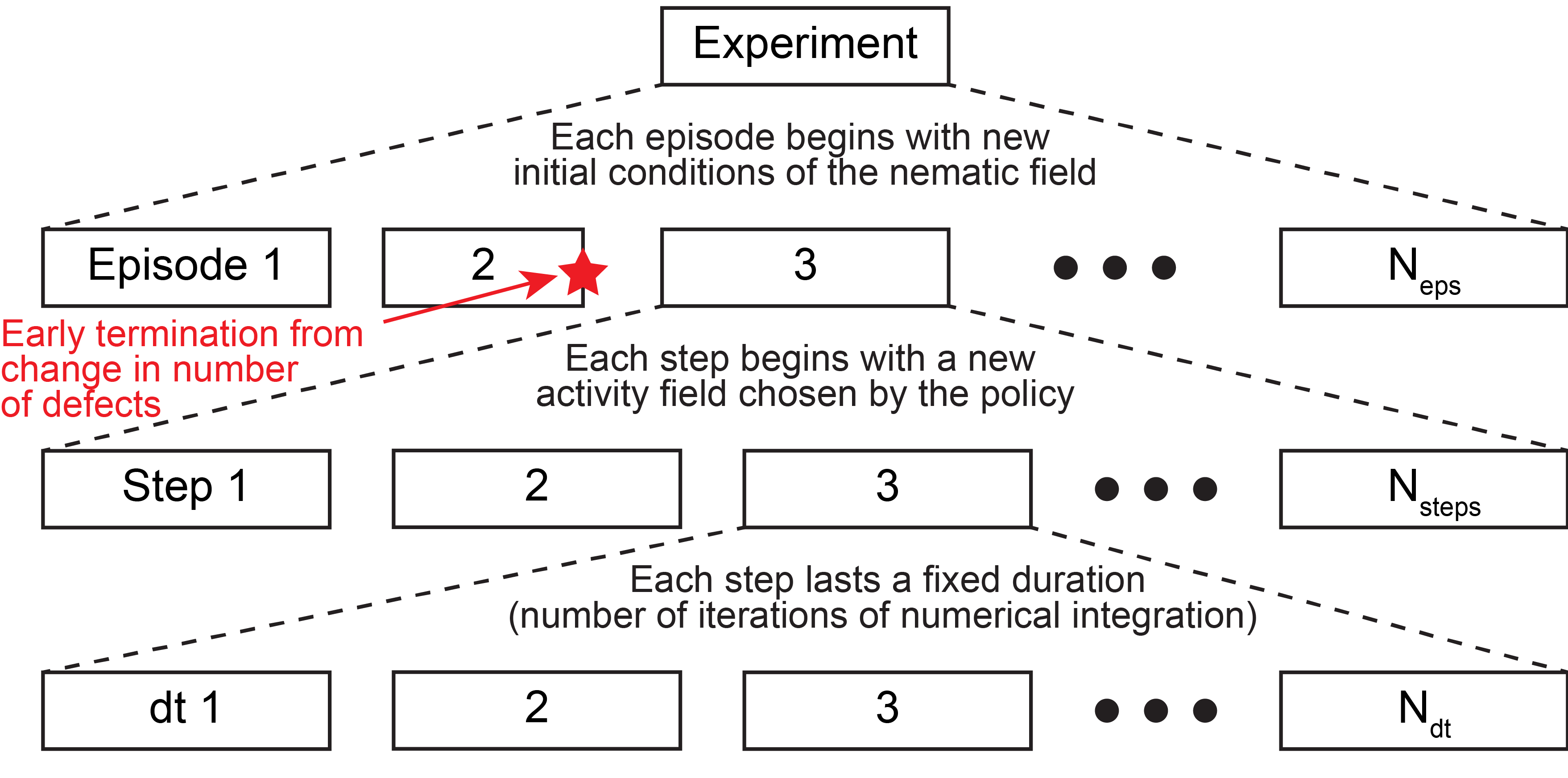}
\caption{Schematic illustration of the structure of an experiment, episode, and step as used in our RL training program. }
\label{TimeDivisions}
\end{center}
\end{figure}
\subsection{States, actions, rewards, and experiment structure}\label{sec:sar}
A RL algorithm produces a so-called policy $\pi_{\boldsymbol{\theta}}(S)$ which is a function that maps from a state $S$ into an action $A$.  Training occurs through many trial-and-error examples in which the policy is incrementally improved via updates to its parameters $\boldsymbol{\theta}$, in the sense that the actions which it learns to choose produce future states that optimize a user-defined reward function.  We demonstrate RL-based control over active nematic defect dynamics through several tasks in this paper, each corresponding to a different choice of states, actions, and reward function.  We describe each choice as we discuss the tasks below.  

For every task we run several experiments using different task parameters or different seeds for the pseudo-random number generator, and the outcome of each experiment is a trained policy function $\pi_{\boldsymbol{\theta}}(S)$.  An experiment is divided into episodes, each of which begins by resetting the active nematic field using some distribution of initial conditions (Figure \ref{TimeDivisions}).  Every episode consists of fixed number $N_\text{steps}$ of steps, although an episode may terminate before this number of steps if the number of defects changes through annihilation or creation.  At every step, the RL agent views the state $S$ and chooses a new action $A$, corresponding to a given configuration of the activity field $\alpha(\mathbf{r})$, which is applied for the duration of the step.  A step consists of $N_{dt}$ iterations of the numerical integrator.  The fixed time interval which elapses during one step is thus $N_{dt} dt $ where $dt$ is the time resolution of the integrator; in simulation units we have $dt = 1$.  The experiment ends when a number $N_\text{eps}$ of episodes is reached or when the wall time of the program exceeds a specified value.  Throughout this paper we set $N_{dt} = 50$, $N_\text{steps} = 75$, and we run experiments for $3$ hours, corresponding to roughly $N_\text{eps} = 150$ episodes per experiment.  

\begin{figure*}[ht!]
\begin{center}
\includegraphics[width=\textwidth]{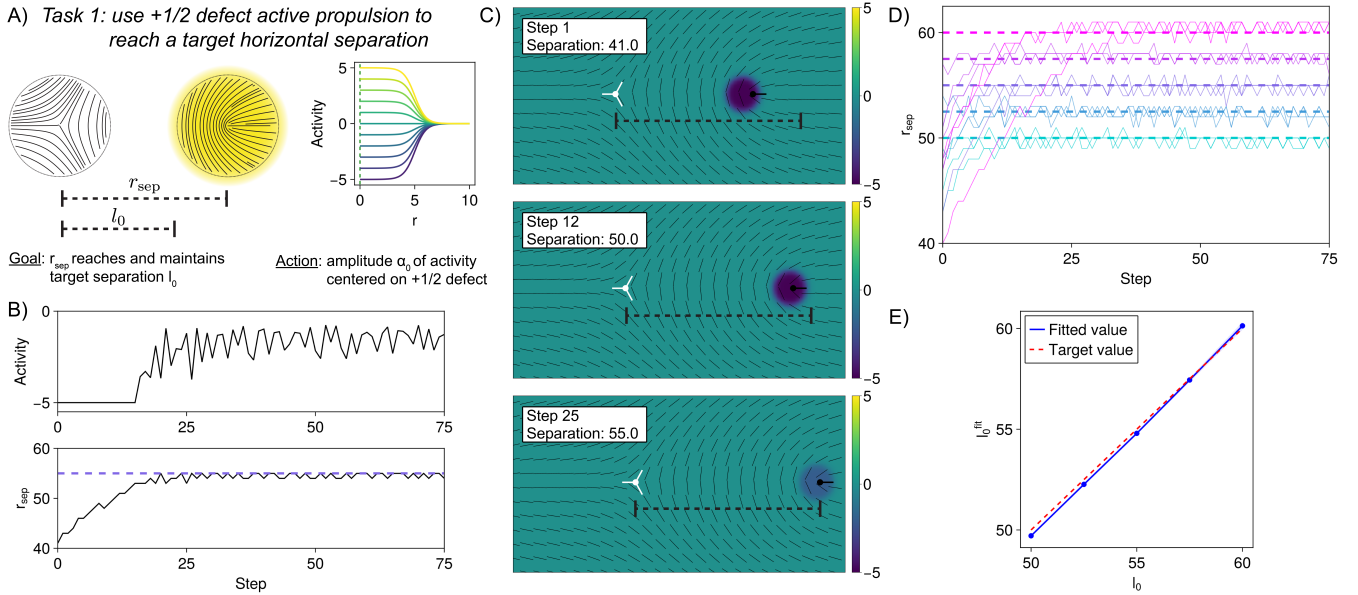}
\caption{Results for task 1.  
A) Schematic illustration of the task, in which a nearly uniform disk of activity is applied to the center of the $+1/2$ defect.  The amplitude of the activity is controlled as an action, and its radial profile for different amplitudes is shown on the right.  The dashed green line denotes the position of the $+1/2$ defect in the activity field at the start of each step.  The goal is to reach a target horizontal separation $l_0$.  
B) Trajectories of the activity amplitude (top) and resulting horizontal separation $r_\text{sep}$ (bottom) with a trained RL policy.  The dashed line represents $l_0 = 55$ for this task.  
C)  Snapshots of the active nematic field at the end of three different steps for the episode in panel B.  The black lines represent the nematic orientation, which are subsbampled so that one line is drawn for every 16 lattice sites.  Color denotes the amplitude of the activity.  The $-1/2$ defect and its orientation are depicted as a white dot and lines, and the $+1/2$ defect and its orientation are depicted as black dot and a line.  The black dashed line represents the target separation $l_0 = 55$.  The 100$\times$100 simulation domain is cropped to 100$\times$50 to ease visibility.
D) Two trajectories of $r_\text{sep}$ with a trained RL policy for five values of $l_0$, shown as different colors.  The values of $l_0$ are shown as dashed lines.  
E) Fitted values of $l_0^\text{fit}$, plotted against the target value of $l_0$.  Small standard deviations are shown as shaded areas.  }
\label{SepPlusResults}
\end{center}
\end{figure*}

\subsection{Reinforcement learning algorithm}\label{sec:rl}
To train the RL policy $\pi_{\boldsymbol{\theta}}(S)$ we use a variant of the actor-critic algorithm \cite{sutton2018reinforcement} called deep deterministic policy gradient (DDPG) \cite{silver2014deterministic, lillicrap2015continuous}, which is suited for continuous actions in deterministic environments.  Four neural networks are used in this approach: two copies of an actor network with parameters $\boldsymbol{\theta}$ and $\boldsymbol{\theta}'$, which implement the policy function, and two copies of a critic network with parameters $\mathbf{w}$ and $\mathbf{w}'$, which estimate the value function $Q_{\mathbf{w}}(S, A)$, i.e., the expected cumulative future reward of choosing action $A$ in state $S$. The main networks, $\mu_{\boldsymbol{\theta}}(A)$ and $Q_{\mathbf{w}}(S, A)$, are updated during training and used to select actions and train the actor, while the target networks track the parameters of the main networks with slow updates to provide stable function estimates for training the critic. Specifically, the target network parameters are updated as $\boldsymbol{\theta}' \leftarrow \rho \boldsymbol{\theta}' + (1-\rho) \boldsymbol{\theta}$ and $\mathbf{w}' \leftarrow \rho \mathbf{w}' + (1-\rho) \mathbf{w}$ for $\rho \lesssim 1$.
This use of target networks improves stability and convergence of learning \cite{lillicrap2015continuous}.

The main actor and critic networks are updated using stochastic gradient descent (SGD) with mini-batches of $N_\text{batch} = 32$ samples from a replay buffer, which stores tuples $(S(s), A(s), R(s), T(s))$ (state, action, reward, and termination flag) collected at every step $s$.  For each experiment, an initial random policy is used for $10 N_\text{batch}$ steps to populate the replay buffer, after which the main actor network $\mu_{\boldsymbol{\theta}}(A)$ is used for action selection, with SGD updates performed every 5 steps.  How the mini-batches are used to update the network parameters is described in Refs.~\citenum{silver2014deterministic, lillicrap2015continuous}.  All neural networks have 2 hidden layers with 32 neurons each, and they are trained using the ADAM optimizer with a learning rate of $0.001$ and a weight norm clip of 1.0.  The discount factor is $\gamma = 0.99$ (see Refs.~\citenum{lillicrap2015continuous, silver2014deterministic, lillicrap2015continuous} for a definition) and the weight transfer factor is $\rho = 0.995$.  

The actions are chosen as $A = \pi_{\boldsymbol{\theta}}(S) + \epsilon$, where  $\epsilon \sim \mathcal{N}(0,0.05)$  is a small Gaussian noise added to the policy output to encourage exploration of the available actions.  The resulting action $A$ is then clipped to the range $[-1,1]$.  State values provided as inputs to the network are normalized to lie approximately within the same range. For each task, the action values are linearly mapped from this range to the corresponding scaled parameters of the activity field $\alpha(\mathbf{r},t)$.  

To numerically integrate the active nematic hydrodynamics, Equation \ref{eqQdyn}, we use a custom Julia implementation of Heun's finite difference method.  The implementation has previously been described and validated in Refs.~\citenum{floyd2023simulating, floyd2024pattern, redford2024motor}.  We note that for our parameterization of the nematic system the unit of length is equal to the equilibrium nematic persistence length (see Appendix) \cite{hemingway2016correlation}.  To train the RL policy, we combine this numerical solver with an implementation of the DDPG algorithm provided by the ReinforcementLearning.jl package \cite{Tian2020Reinforcement}.

\section{Results}
To demonstrate the feasibility of using RL to control active nematic defect interactions through spatiotemporal activity fields, we show here several test cases of varying difficulty.  

\subsection{Translating a $+1/2$ defect}

The first task we consider is effectively one-dimensional, in which we nucleate a pair of defects that are vertically aligned in the $100 \times 100$ periodic domain and horizontally separated by a random offset $l_\text{init}$ selected uniformly from the range $[37.5, 62.5]$.  The goal is to move the defects so that they are horizontally separated by an amount $l_0$ (Figure \ref{SepPlusResults}A).  As a first demonstration, in this task we are not imposing a virtual dynamics but instead imposing a static property that the defects should exhibit.  To do this we apply a disk of nearly uniform activity centered on the $+1/2$ defect, and we leverage the fact that such a defect propels with a velocity vector parallel to its orientation vector and approximately proportional to the local activity: $\dot{\mathbf{p}}_+ \sim - \alpha \hat{\mathbf{e}}$ \cite{giomi2014defect, shankar2018defect, shankar2024design}.  

The state used in the RL algorithm at step $s$ is ${S(s) = (r_\text{sep}(s) - l_0) / 50}$ (where $50$ is a rough scale factor), and the action is the amplitude $\alpha_0$ of the activity profile scaled to the range $[-5,5]$.  The shape of the activity profile is an azimuthally symmetric function centered on $\mathbf{p}_+$ whose radial dependence is
\begin{equation}
    \alpha(r;\alpha_0,c,m) = \frac{\alpha_0}{2}\left(1 - \tanh\left(\frac{r - c}{m} \right) \right), \label{tanhprof}
\end{equation}
where, for this task, $c = 5$ is a cutoff parameter and $m = 1$ the width of the logistic profile.  The reward for this task is computed at step $s$ as $R(s) = - |r_\text{sep}(s+1) - l_0|$.  

For a trained policy in which $l_0 = 55$, we observe a typical episode as in Figures \ref{SepPlusResults}B and C.  The policy applies a strong negative activity field to the $+1/2$ defect to cause it to propel away from the $-1/2$ defect.  Upon reaching the desired separation of $55$ lattice units the applied activity then weakens in magnitude and, in a feedback control-like manner, nudges the $+1/2$ defect further backward whenever the $-1/2$ defect moves toward it due to the attractive elastic interaction between them.  This maintains the separation at $r_\text{sep} = 55$ for the remainder of the episode.  

\begin{figure*}[ht!]
\begin{center}
\includegraphics[width=\textwidth]{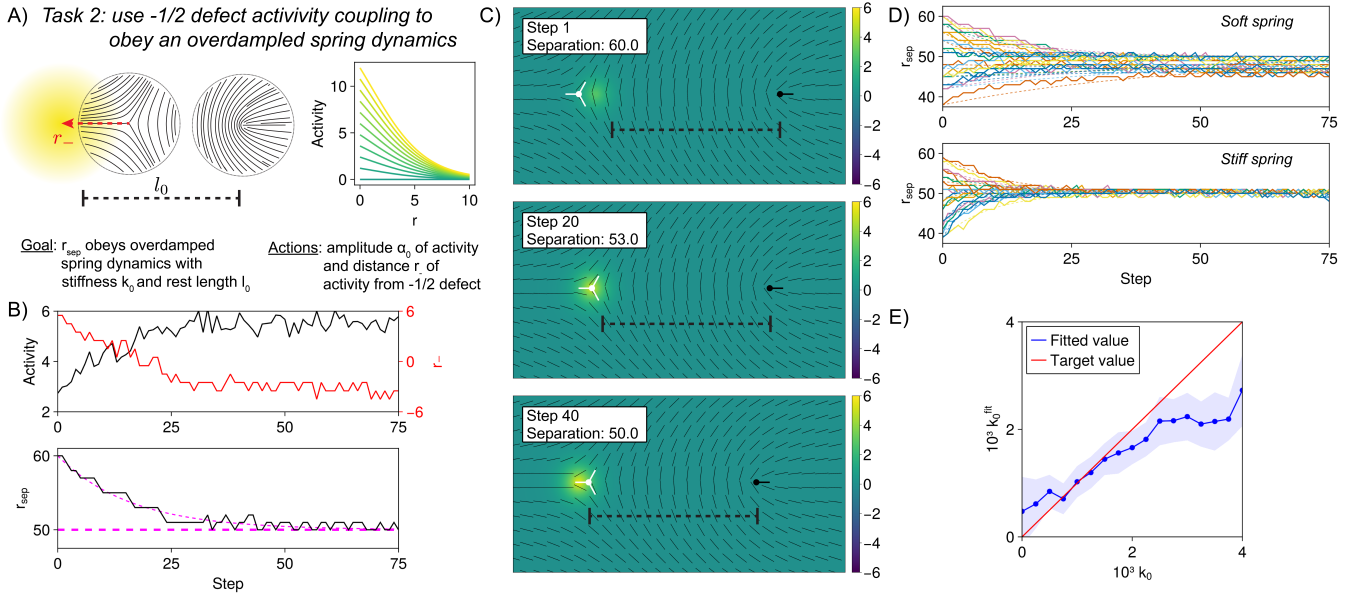}
\caption{Results for task 2.  
A) Schematic illustration of the task, in which an inhomogeneous activity is field applied near the $-1/2$ defect.  The position of the activity field and its amplitude are controlled as actions, and its radial profile for different amplitudes is shown on the right.  The goal is for $r_\text{sep}$ to obey the dynamics of an overdamped spring with rest length $l_0 = 50$ and varying stiffness $k_0$.  
B) \textit{Top}:  Trajectories of the activity amplitude (black) and horizontal distance from the $-1/2$ defect (red) for a trained RL policy. \textit{Bottom}: The resulting horizontal separation $r_\text{sep}$ trajectory.  The thick dashed line on the bottom represents $l_0 = 50$, the target stiffness is $k_0 = 0.0015$, and and the thin dashed line denotes an exponential fit with $k^\text{fit}_0 = 0.00124$.  
C)  Snapshots of the active nematic field at the end of three different steps for the episode in panel B.  See Figure \ref{SepPlusResults} for a description.  The black dashed line represents the target separation $l_0 = 50$.  
D) \textit{Top}: A set of 20 trajectories of $r_\text{sep}$ with a trained RL policy for $k_0 = 0.001$, with exponential fits shown as dashed lines.  \textit{Bottom}: Same as top, but with $k_0 = 0.004$.
E) Fitted values $k_0^\text{fit}$ plotted against the target value of $k_0$.  Standard deviations are shown as shaded areas. For this plot we exclude episodes which randomly start within $2$ of $r_\text{sep} = 50$ to focus on trajectories in which $r_\text{sep}$ changes appreciably during the episode.}
\label{SepMinResults}
\end{center}
\end{figure*}

We trained RL policies for a range of values of $l_0$ and observed that in each case the algorithm converged to a policy in which the target separation was reached by the end of each episode (Figure \ref{SepPlusResults}D).  We fit the learned $l_0^\text{fit}$ by taking the average of $r_\text{sep}$ over the last 25 steps of each episode, over the last 40 episodes of each experiments, and over 10 experiments using different random initial seeds.  The learned $l_0^\text{fit}$ matches the target $l_0$ (Figure \ref{SepPlusResults}E) in each case.  Thus, RL can learn how to adjust the activity amplitude as a function of defect position to reach a target separation, for a simple physical set-up in which the defects are aligned and move primarily due to the activity-induced propulsion of the $+1/2$ defect.

\subsection{Translating a $-1/2$ defect}
We next consider a more complicated task, in which the horizontal defect separation $r_\text{sep}$ should not just maintain a user-defined static value but should evolve under a user-defined dynamics.  We specify an overdamped spring dynamics, in which
\begin{equation}
\dot{r}^*_\text{sep} = -k_0(r_\text{sep} - l_0) \label{eqsprdyn}
\end{equation}
where the asterisk denotes that this is the target dynamics of $r_\text{sep}$.  For step of duration $\Delta t = N_{dt} dt$ (cf. Figure \ref{TimeDivisions}), we use the finite-difference approximation to the change in $r_\text{sep}$ from the target dynamics at step $s$,
\begin{equation}
    r^*_\text{sep}(s+1) = r_\text{sep}(s) - \Delta t k_0(r_\text{sep}(s) - l_0),
\end{equation}
to form the reward function at this step 
\begin{equation}
    R(s) = - |r^*_\text{sep}(s+1) - r_\text{sep}(s+1)| \label{eqrewfd}
\end{equation}
where $r_\text{sep}(s+1)$ is the actual separation obtained by evolving the dynamics under the activity field chosen at step $s$.  As the RL program learns to better mimic the prescribed dynamics the reward approaches $0$ from below.  The state observed by the program and the episode initialization procedure are the same as in the previous task.  

\begin{figure*}[ht!]
\begin{center}
\includegraphics[width=\textwidth]{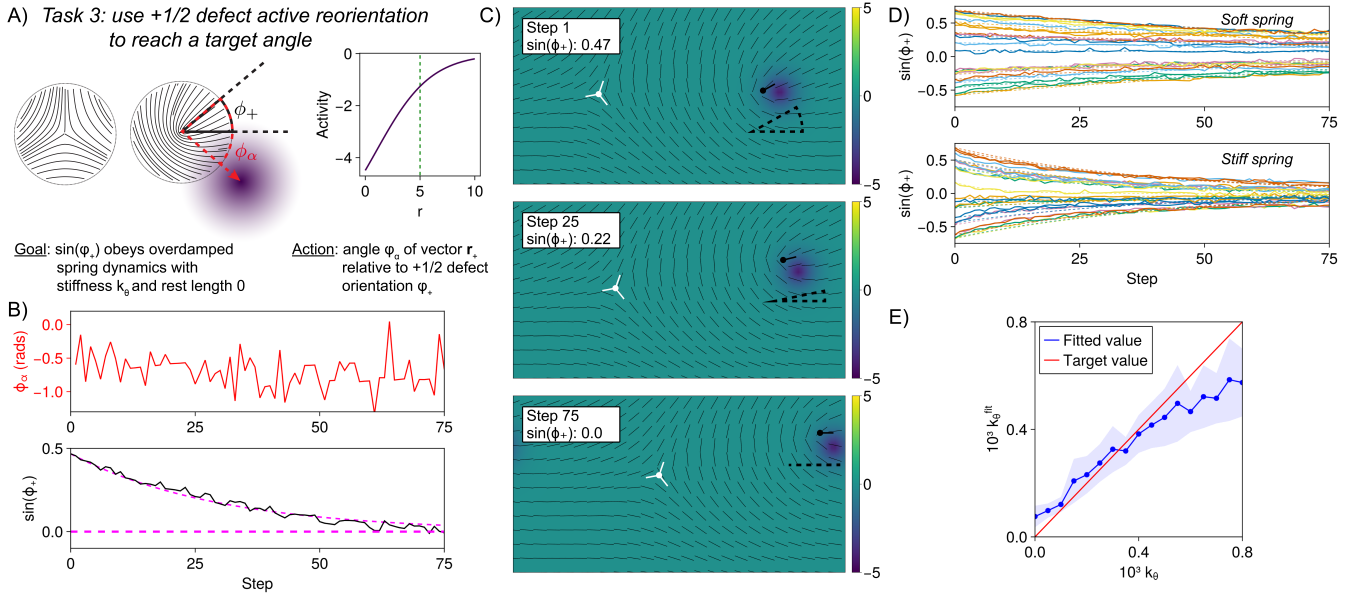}
\caption{Results for task 3.  A) Schematic illustration of the task, in which an inhomogeneous activity is field applied near the $+1/2$ defect.  The negative angle $\phi_\alpha$ is indicated.  This angle is controlled as an action, and the activity field's radial profile is shown on the right.  The dashed green line denotes the position of the $+1/2$ defect in the activity field at the start of each step.  The goal is for $\sin(\phi_+)$ to obey the dynamics of an overdamped spring with rest value $0$ and varying stiffness $k_\theta$. 
B) \textit{Top}:  Trajectory of the angular position $\phi_\alpha$ of the activity field for a trained RL policy. \textit{Bottom}: The resulting trajectory for orientation of the $+1/2$ defect $\sin(\phi_+)$.  The thick dashed line on the bottom represents $\sin(\phi_+) = 0$, the target stiffness is $k_\theta = 0.0007$, and and the thin dashed line denotes an exponential fit with $k^\text{fit}_\theta = 0.00067$.  
C)  Snapshots of the active nematic field at the end of three different steps for the episode in panel B.  See Figure \ref{SepPlusResults} for a description.  The black dashed wedge represents the current orientation $\sin(\phi_+)$ which approaches the target value of $0$.  
D) \textit{Top}: A set of 20 trajectories of $r_\text{sep}$ with a trained RL policy for $k_\theta = 0.00025$, with exponential fits shown as dashed lines.  \textit{Bottom}: Same as top, but with $k_\theta = 0.0008$.
E) Fitted values $k_\theta^\text{fit}$ plotted against the target value of $k_\theta$.  For this plot we exclude episodes which randomly start within $0.05$ of $\sin(\phi_+) = 0$ to focus on trajectories in which $\sin(\phi_+)$ changes appreciably during the episode. }
\label{OprPlusResults}
\end{center}
\end{figure*}

We train the RL program to achieve these target dynamics by leveraging a recently demonstrated modality of defect motion under activity \textit{gradients}, rather than the modality of $+1/2$ defect propulsion under constant activity used in the previous task \cite{shankar2024design}.  $-1/2$ defects have been shown to couple to second-order and higher gradients in $\alpha(\mathbf{r})$, and, using this fact, we parameterize the activity field in the vicinity of the $-1/2$ defect as $\alpha(r;\alpha_0,1,5)$ in Equation \ref{tanhprof}, which has non-zero gradients of all orders.  We allow the RL program to adjust the amplitude $\alpha_0$ of the activity field (in the range $[0,12]$) and the offset $r_-$ between the $-1/2$ defect and the center of the activity field (in the range $[-10,10]$); see Figures \ref{DefectGeometry} and \ref{SepMinResults}A.  

We fix $l_0 = 50$ in Equation \ref{eqsprdyn} throughout and vary $k_0$, the overdamped spring constant.  For a trained policy in which $k_0 = 0.0015$ we observe typical trajectories as in Figures \ref{SepMinResults}B and C.  The policy simultaneously adjusts the amplitude and offset of the activity profile relative to the $-1/2$ defect so that the defect separation $r_\text{sep}$ gradually approaches the ``rest length'' of $l_0 = 50$, in an exponential decay determined by the stiffness $k_0$.  We fit an exponential curve to this decay and obtain $k_0^\text{fit} = 0.00124$, in decent agreement with the target value.  Viewing several episodes for policies trained on $k_0 = 0.001$ and $k_0 = 0.004$, it is apparent that the target stiffness indeed manifests as a different rate of approach of $r_\text{sep}$ toward $l_0$ (Figure \ref{SepMinResults}D).   

We consider a range of $k_0$ values and average $k_0^\text{fit}$ over the last $40$ episodes of each experiment and over $10$ experiments using different random initial seeds, with results shown in Figure \ref{SepMinResults}E.  We observe good agreement between $k_0$ and $k_0^\text{fit}$, albeit with deviations at high values of $k_0$.  At these values the policy has difficulty pulling the defects faster than the nematic material timescales allow.  Despite these limitations, the generally good agreement between $k_0$ and $k_0^\text{fit}$ indicates that the RL program can utilize a range of physical effects (including coupling to gradients of activity) to pull defects, and they can mimic target dynamics rather than just target static configurations.

\subsection{Rotating a $+1/2$ defect}
Finally, we consider the task reorienting the $+1/2$ defect by coupling its orientation vector to gradients in the activity field. Describing the defect's orientation by $\zeta_+ \equiv \sin(\phi_+)$, we specify the dynamical law
\begin{equation}
    \dot{\zeta}_+^* = - k_\theta \zeta_+, \label{eqzetadot}
\end{equation}
corresponding to a overdamped spring with rest length 0 acting on the variable $\zeta_+$.  We use $\sin(\phi_+)$ instead of $\phi_+$ to handle the discontinuity at $\phi_+ = 0$ and $\phi_+ = 2\pi$.  As in the previous task we use the finite-difference approximation for $\zeta^*_+(s+1)$ to compute the reward $R(s) = - |\zeta_+^*(s+1) - \zeta_+(s+1)|$.  The state observed by the RL program is $S(s) = \zeta_+(s)$, and its action is the angle $\phi_\alpha$, depicted in Figures \ref{DefectGeometry} and \ref{OprPlusResults}A, between $\hat{\mathbf{e}}$ and the center of the activity profile $\alpha(r;-7.5,5,1)$ which is centered $r_+ = 5$ lattice units away from the defect.  We draw initial conditions by taking a pair of defects vertically aligned and horizontally separated by $50$ lattice points, and then rotating the nematic director everywhere by a random angle chosen uniformly from the range $[-0.4, 0.4]$ rads.

We vary the stiffness $k_\theta$ in Equation \ref{eqzetadot} over a range of values.  For a typical trained policy in which $k_\theta = 0.0007$ we observe trajectories as in Figures \ref{OprPlusResults}B and C.  The policy fluctuates the angle $\phi_\alpha$ around a mean value to cause the $+1/2$ defect to rotate at a desired exponential rate toward $\phi_+ = \zeta_+ = 0$.  We fit an exponential curve to this decay, obtaining $k_\theta^\text{fit} = 0.0067$.  In Figure \ref{OprPlusResults}D we show several episodes for $k_\theta = 0.00025$ and $k_\theta = 0.0007$, indicating a clear difference in the learned decay rate of $\zeta_+$ toward 0.  
Fitting $k_\theta^\text{fit}$ for the last 40 episodes of each experiment and over 10 experiments with different random initial seeds, we observe good agreement with the target value of $k_\theta$ (Figure \ref{OprPlusResults}E).  This indicates that RL can mimic dynamical laws governing both the positional and orientational dynamics of active nematics. 

% \subsection{Complex hybrid interaction task}
% TBD

\section{Discussion}
We have explored a computational strategy for guiding the dynamics of active nematic defects using spatiotemporal activity fields $\alpha(\mathbf{r},t)$ through closed-loop control policies learned by RL. We demonstrated a proof of principle by employing RL to control active nematic defect dynamics in several simple examples, leaving more complex tasks for future work. RL offers a model-free approach for learning feedback control policies, which implement a desired dynamical interaction law, via trial-and-error exploration of the action and reward space. This approach presents a practically viable alternative to optimal control methods \cite{norton2020optimal, ghosh2024achieving} or human-designed techniques \cite{shankar2024design}, which require accurate model specification. Additionally, recent work suggests that RL can offer qualitative advantages over optimal control by learning to optimize more effective reward functions \cite{song2023reaching}. Future research could directly compare RL and optimal control in terms of the robustness of their policies for controlling active nematics.

The scientific merit of demonstrating control over active nematics via RL is at least two-fold. First, as mentioned earlier, RL offers a practical method for engineering soft active matter systems with minimal reliance on accurate model specification, although it does present several technical challenges. When deploying RL to control a new system, careful attention must be given to parameterizing the control fields and designing the reward functions in a way that allows the algorithm to efficiently learn to solve the abstract optimization problem. Like other machine learning applications, this process also involves fine-tuning various hyperparameters such as learning rates, neural network architectures, etc. Additionally, reliable state observations are important (although there exist stochastic variants of RL using uncertain state measurements), and accurately measuring nematic field configurations in experiments requires considerable technical attention \cite{li2024machine, tran2024deep}. Despite these challenges, RL has a proven track record of solving highly complex problems, often surpassing human capabilities \cite{silver2016mastering, reddy2018glider}, making its application to controlling active matter systems quite promising.

Second, the ability of RL programs to successfully manipulate the dynamics of active nematic defects using very low-dimensional state and action spaces suggests that an effective low-dimensional description is sufficiently accurate to capture the defects' dynamics. This finding supports the theoretical arguments presented in Refs.~\citenum{shankar2018defect} and \citenum{shankar2024design}. Moreover, in our previous work, we demonstrated that imperfect, spatiotemporally local feedback signals are sufficient for learning open-loop control policies to guide a range of non-equilibrium dynamical systems \cite{floyd2024learning}. The adequacy of these imperfect feedback protocols, which rely only on coarse projections of the entire nematic field, makes biologically-implemented control over defects more plausible \cite{maroudas2021topological}.

\section*{Acknowledgments}
We wish to thank Alexandra Lamtyugina, Luca Scharrer, Suraj Shankar, Grant Rotskoff, Michael Hagan, Saptorshi Ghosh, and Aparna Baskaran for helpful discussions.  This work was supported by DOE BES Grant DE-SC00197, National Science Foundation (NSF) award MCB-2201235, and the University of Chicago Materials Research Science and Engineering Center, which is funded by the NSF under award number DMR-2011854. The authors acknowledge the University of Chicago’s Research Computing Center for computing resources.

\appendix
\section*{Appendix: Active nematic equations of motion} 
Active nematic fluids are described by an order parameter $\mathbf{Q}$, which is a symmetric and traceless tensor:
\begin{equation}
    \mathbf{Q} = q\left(\hat{\mathbf{n}}\hat{\mathbf{n}} - \frac{1}{d}\mathbf{I}\right),
\end{equation}
where $\hat{\mathbf{n}}$ is a unit director, $q$ measures the degree of polarization, $d$ is the dimensionality, and $\mathbf{I}$ is the identity tensor. Our treatment follows the experimentally-motivated model in Refs.~\citenum{zhang2021spatiotemporal, colen2021machine, redford2024motor}, which uses $d = 3$ but projects the field onto a 2D plane to which the nematic is spatially confined, while allowing it to rotate out of this plane. We note that this differs slightly from the strictly 2D treatment used in several theoretical works such as Refs.~\citenum{shankar2018defect, shankar2024design}. 

The tensor $\mathbf{Q}$ couples to a flow field $\mathbf{v}$ and spontaneously seeks to minimize its local free energy. We consider the overdamped limit and the limit of high substrate friction (as in Refs.~\citenum{floyd2024learning, shankar2024design}), yielding the following equations of motion:
\begin{align}
    \partial_t Q_{ij} &= S_{ij}(\mathbf{v}) + \Gamma_H H_{ij}, \label{eqQdyn} \\
    v_i &= \gamma_{v}^{-1} \partial_k\left(\sigma^a_{ik}(\mathbf{Q}) + \sigma^E_{ik}(\mathbf{Q}) \right). \label{eqv}
\end{align}

Here, $H_{ij}$ is the symmetric and traceless part of $-\frac{\delta F}{\delta Q_{ij}}$ where $F$ is the free energy functional (see below), $S_{ij}$ is a flow-coupling term, and $\gamma_v$ is a coefficient of friction. In this overdamped limit, $\mathbf{v}$ is an instantaneous function of $\mathbf{Q}$, so Equation~\eqref{eqQdyn} is closed in $\mathbf{Q}$. This simplification allows for computational efficiency which is advantageous in training RL programs. The active stress tensor is given by \cite{hatwalne2004rheology, marchetti2013hydrodynamics}
\begin{equation}
    \sigma^a_{ij} = -\alpha Q_{ij},
\end{equation}
and the Ericksen stress tensor is given by \cite{beris1994thermodynamics}
\begin{align}
    \sigma^E_{ij} &= f\delta_{ij} -\xi H_{ik}\left(Q_{kj}+\frac{1}{3}\delta_{kj} \right) - \xi\left(Q_{ik}+\frac{1}{3}\delta_{ik} \right)H_{kj} \nonumber \\
    &\quad + 2\xi \left(Q_{ij}+\frac{1}{3}\delta_{ij}\right)H_{kl}Q_{kl} - \partial_j Q_{kl}\frac{\delta F}{\delta \partial_i Q_{kl}} \nonumber \\
    &\quad + Q_{ik}H_{kj} - H_{ik}Q_{kj}.
\end{align}

The Landau-de Gennes free energy is:
\begin{equation}
    F = \int d\mathbf{r} \, f(\mathbf{r}),
\end{equation}
where
\begin{align}
    f &= \frac{A_0}{2}\left(1-\frac{U}{3}\right)\text{Tr}\left(\mathbf{Q}^2\right) - \frac{A_0 U}{3}\text{Tr}\left(\mathbf{Q}^3\right) \nonumber \\
    &\quad + \frac{A_0 U}{4}\left(\text{Tr}\left(\mathbf{Q}^2\right)\right)^2 + \frac{L}{2}\left(\partial_kQ_{lm}\right)^2.
\end{align}
The flow coupling term is 
\begin{align}
    S_{ij}(\mathbf{v}) =& -v_k \partial_k  Q_{ij} + \Phi_{ik} Q^+_{kj} +Q^+_{ik}\Phi_{kj} \nonumber - 2\xi Q^+_{ij} \left(Q_{kl}\partial_k v_l\right),
\end{align}
where
\begin{equation}
    Q^+_{ij} = Q_{ij} + \frac{1}{3}\delta_{ij},
\end{equation}
\begin{equation}
    \Psi_{ij} = \frac{1}{2}\left(\partial_i v_j + \partial_j v_i \right),
\end{equation}
\begin{equation}
    \Omega_{ij} = \frac{1}{2}\left(\partial_i v_j - \partial_j v_i \right),
\end{equation}
and
\begin{equation}
    \Phi_{ij} = \xi \Psi_{ij} - \Omega_{ij}.
\end{equation}
In these equations, $\xi$, $A_0$, $U$, $\Gamma_H$, $L$, $\gamma_v$ are parameters whose meanings are described in  Refs.~ \citenum{zhang2021spatiotemporal, colen2021machine, redford2024motor}, to which we refer for additional details on this model.  Throughout the paper we use $\xi = 0.7$, $A_0 = 0.1$, $U = 3.5$, $\Gamma_H = 1.5$, $L=0.1$, and $\gamma_v = 10$ (all in simulation units).  For these choices the equilibrium nematic persistence length $\sqrt{L/A_0}$ is the same as one length unit \cite{hemingway2016correlation}.

\end{twocolumngrid}

\bibliographystyle{unsrt}

\end{document}